\documentclass[twocolumn]{jpsj3}
\usepackage{color}

\title{
Temperature-Dependent Cycloidal Magnetic Structure in GdRu$_{2}$Al$_{10}$ Studied by Resonant X-ray Diffraction
}

\author{Takeshi Matsumura$^{1,2,3,4}$, Takayoshi Yamamoto$^{1}$, Hiroshi Tanida$^{1}$, and Masafumi Sera$^{1,2}$}

\inst{$^{1}$Department of Quantum Matter, AdSM, Hiroshima University, Higashi-Hiroshima, Hiroshima 739-8530, Japan \\
$^{2}$Institute for Advanced Materials Research, Hiroshima University, Higashi-Hiroshima, Hiroshima 739-8530, Japan \\
$^{3}$Chirality Research Center (CResCent), Hiroshima University, Higashi-Hiroshima, Hiroshima 739-8526, Japan\\
$^{4}$Center for Emergent Condensed Matter Physics, Hiroshima University, Higashi-Hiroshima, Hiroshima 739-8526, Japan\\
}

\abst{
We have performed resonant X-ray diffraction experiments on the antiferromagnet GdRu$_{2}$Al$_{10}$ and have clarified that the magnetic structure in the ordered state is cycloidal with the moments lying in the $bc$ plane and propagating along the $b$ axis. 
The propagation vector shows a similar temperature dependence to the magnetic order parameter, which can be interpreted as being associated with the gap opening in the conduction band and the resultant change in the magnetic exchange interaction. 
Although the $S=7/2$ state of Gd is almost isotropic, the moments show slight preferential ordering along the $b$ axis. 
The $c$ axis component in the cycloid develops with decreasing temperature through a tiny transition in the ordered phase.  
We also show that the scattering involves the $\sigma$-$\sigma'$ process, which is forbidden in normal $E1$-$E1$ resonance of magnetic dipole origin. We discuss the possibility of the $E1$-$E2$ resonance originating from a toroidal moment due to the lack of inversion symmetry at the Gd site. 
The spin-flop transition in a magnetic field is also described in detail. 
 }

\begin{document}

\maketitle
\section{Introduction}
Hybridization between localized and itinerant electrons gives rise to a rich variety of electronic states through competition between the Kondo effect and the Ruderman--Kittel--Kasuya--Yosida (RKKY) magnetic exchange interaction.\cite{Otsuki09} 
In $f$-electron systems, the former leads to a nonmagnetic heavy-fermion state, or in some cases, a Kondo semiconducting state, whereas the latter preferentially leads to a magnetic ordered state. 
Recently, a new type of Kondo semiconductor system, Ce$T_2$Al$_{10}$ ($T$=Ru and Os), has been attracting continuous interest because of its unconventional nature of a long-range magnetic order coexisting with the Kondo effect due to strong $c$-$f$ hybridization.\cite{Muro09,Nishioka09,Strydom09,Tanida10a,Tanida10b,Robert10,Kimura11,Kondo11,Sakoda11,Robert12} 
The most prominent feature is the high transition temperatures; $T_{\text{N}}=27.3$ K for CeRu$_2$Al$_{10}$ and 
28.7 K for CeOs$_2$Al$_{10}$, which are higher than $T_{\text{N}}=17.5$ K for GdRu$_2$Al$_{10}$ and cannot be understood as being caused by the normal RKKY exchange interaction. 
In spite of the extensive studies, the true mechanism of the ordering phenomenon has not yet been clarified. 

Ce$T_2$Al$_{10}$ is also anomalous from the viewpoint of its magnetic propagation vector. 
The magnetic moments in Ce$T_2$Al$_{10}$ order along the $c$ axis with the propagation vector $\mib{q}=(0,1,0)$.\cite{Khalyavin10,Khalyavin13,Kobayashi14} 
In isostructural NdFe$_2$Al$_{10}$ and SmRu$_2$Al$_{10}$, on the other hand, the magnetic structure is described by $\mib{q}=(0,3/4,0)$ and its third harmonic of $(0, 1/4, 0)$, reflecting the squaring up.\cite{Robert14,Takai15} 
The high-temperature phase of SmRu$_2$Al$_{10}$ is an incommensurate phase with $\mib{q}=(0,0.759,0)$. 
There is also a case of $\mib{q}\simeq (0, 0.798, 0)$ as in TbRu$_2$Al$_{10}$.\cite{Reehuis03} 
Therefore, $\mib{q}=(0, 0.75 - 0.8, 0)$ seems to be a common propagation vector corresponding to the RKKY interaction in the $RT_2$Al$_{10}$ series of compounds. 
Although $\mib{q}=(0,1,0)$ for $R$=Ce reflects some common characteristic of this crystal structure in the sense that $\mib{q}$ lies along the $b$ axis, the slight difference from other isostructural compounds may have some association with the $c$-$f$ hybridization effect.

GdRu$_2$Al$_{10}$ shows an antiferromagnetic order at $T_{\text{N1}}=17.5$ K, followed by another weak transition at $T_{\text{N2}}=16.5$ K.\cite{Sera13a} 
The magnetic susceptibility $\chi(T)$ and magnetization $M(H)$ of GdRu$_2$Al$_{10}$ can well be explained by a mean-field model calculation and can be understood as a simple antiferromagnet of almost isotropic Gd$^{3+}$ with $S=7/2$. 
Since $\chi(T)$ shows a cusp anomaly for $H\parallel b$ and $c$, the ordered moments are expected to be in the $bc$ plane. 
This is also consistent with the spin-flop transition observed for $H\parallel b$ and $c$. 
However, the actual magnetic structure has not yet been clarified. 

The purpose of the present study is to investigate the magnetic structure of GdRu$_2$Al$_{10}$ by resonant X-ray diffraction. 
Since Gd is a strong neutron-absorbing element, the magnetic structure has not been determined yet by neutron diffraction. 
Resonant X-ray diffraction is more suitable in this study. 
The high space resolution obtained by using a synchrotron X-ray beam is also an advantage, which has been utilized in this study to measure the shift of the peak position with temperature.

\section{Experiment}
Single crystals of GdRu$_2$Al$_{10}$ were prepared by an Al flux method. 
The magnetic susceptibility and magnetization of this sample have already been reported in Ref.~\citen{Sera13a}. 
A resonant X-ray diffraction experiment was performed at BL-3A of the Photon Factory, High Energy Accelerator Research Organization (KEK). Two samples were prepared for the experiment, one with a (100) surface and the other with a (010) surface, which were both mirror-polished. 
A magnetic field was applied using a vertical field 8 T superconducting cryomagnet. We used X-ray energies near the $L_3$ absorption edge of Gd. Polarization analysis was performed using a Cu-220 reflection, where the scattering angle was $84.04^{\circ}$ at 7.246 keV. 
We also used a diamond phase retarder system installed at BL-3A, which enabled us to tune the horizontally polarized incident beam to a circularly polarized beam.

\section{Results and Analysis}
\subsection{Temperature-dependent propagation vector}
\begin{figure}[t]
\begin{center}
\includegraphics[width=7.5cm]{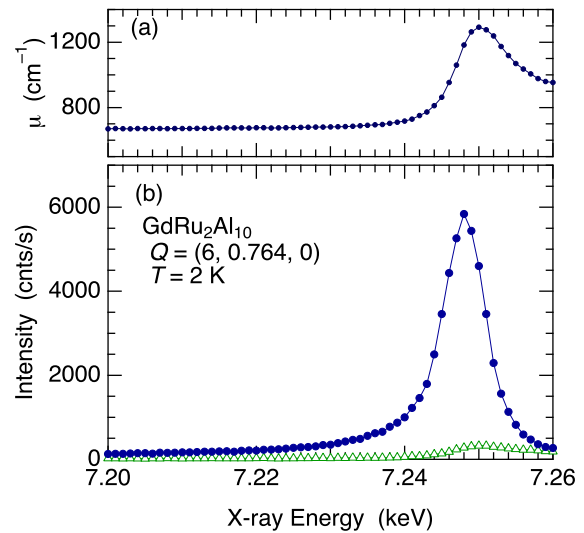}
\end{center}
\caption{(Color online) (a) Absorption coefficient obtained from the fluorescence spectrum. 
(b) X-ray energy dependence of the magnetic Bragg peak at $\mib{Q}=(6, 0.764, 0)$ at 2 K without polarization analysis. 
The triangles represent the background. }
\label{fig:Escan1}
\end{figure}

In the reciprocal lattice scan at the lowest temperature of 2 K, we found clear diffraction peaks at incommensurate positions corresponding to $\mib{Q} = \mib{q} + \mib{\tau}$, where $\mib{q}=(0, 0.764, 0)$ and $\mib{\tau}$ is the reciprocal lattice vector of the fundamental lattice. 
No higher harmonic peaks were detected, indicating that the magnetic structure is described by a single $\mib{q}$ component. 
In Fig.~\ref{fig:Escan1}, we show the energy dependence of the Bragg peak at $\mib{Q} = (6, 0.764, 0) = \mib{q} + \mib{\tau}_{600}$. 
The intensity shows a resonant enhancement at 7.248 keV, corresponding to the $E1$-$E1$ ($2p\leftrightarrow 5d$) resonance. 
No significant change in the fundamental lattice reflections was detected within the accuracy of the present experiment, 
indicating that the magnetostriction, i.e., the coupling with the lattice, is very small. 

The very high count rate of 6000 cps reflects the large ordered moment of Gd$^{3+}$ with $S=7/2$, resulting in a large exchange splitting in the $5d$ state of Gd. 
This enabled us, as described in the following, to investigate the scattering process in detail by performing polarization analysis and by using a phase retarder device, which both lead to a significant reduction of the count rate but nevertheless a reasonable signal intensity remains. 

\begin{figure}[t]
\begin{center}
\includegraphics[width=8cm]{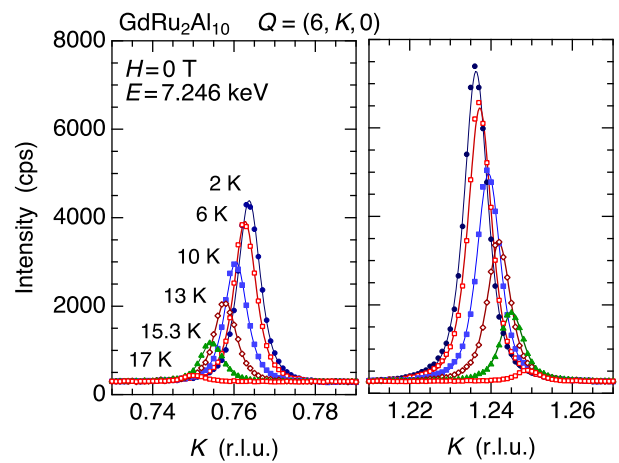}
\end{center}
\caption{(Color online) Scans of the resonant intensity at resonance in the reciprocal space along $\mib{Q}=(6, K, 0)$ 
at several temperatures without polarization analysis. The peaks in the left panel correspond to $\mib{\tau}_{600} + \mib{q}$ and those in the right correspond to $\mib{\tau}_{620} -\mib{q} $. }
\label{fig:KscanTdep}
\end{figure}

\begin{figure}[t]
\begin{center}
\includegraphics[width=7.5cm]{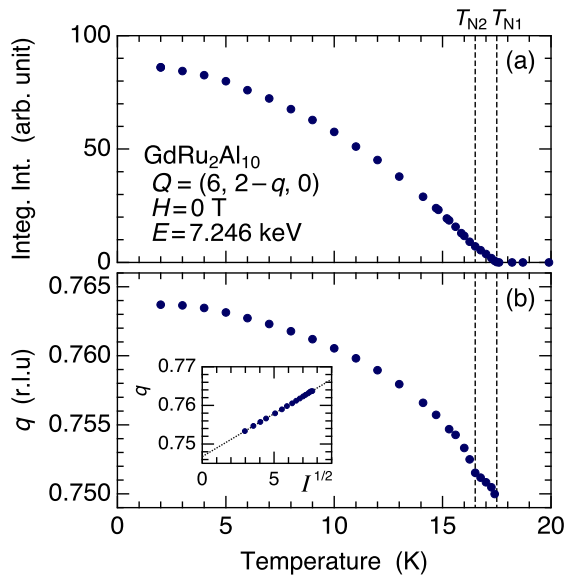}
\end{center}
\caption{(Color online) (a) Temperature dependence of the integrated intensity of the $\mib{\tau}_{620} -\mib{q}$ peak in Fig.~\ref{fig:KscanTdep}. The vertical dashed lines show the phase boundaries observed in the $\chi(T)$ measurement. 
(b) Temperature dependence of the $q$ value obtained from the reciprocal scans of the $\mib{\tau}_{600} +\mib{q}$ and $\mib{\tau}_{620} -\mib{q}$ peaks in Fig.~\ref{fig:KscanTdep}. The inset shows a plot of the $q$ value as a function of the square root of the intensity. }
\label{fig:TdepQInt}
\end{figure}

Figure \ref{fig:KscanTdep} shows the reciprocal scans of the resonance intensity along the $(6, K, 0)$ line at several temperatures. 
Since this peak disappears above $T_{\text{N1}}=17.5$ K, the scattering can be ascribed to magnetic origin, i.e., the $E1$ resonance is associated with the exchange splitting in the $5d$ state of Gd, which is induced by the magnetic moment in the $4f$ orbital.\cite{Hannon88} 
In addition, it is intriguing that the peak position continuously shifts with the temperature. Since the positions of the Bragg peaks of (6, 0, 0) and (6, 2, 0) do not change at all, this shift in the magnetic Bragg peak directly shows that the periodicity of the incommensurate magnetic structure changes with the temperature. 
The temperature dependences of the integrated intensity and the $q$ value obtained from this measurement are shown in Fig.~\ref{fig:TdepQInt}. 
Interestingly, the $q$ value seems to start from nearly 0.75 just below $T_{\text{N1}}=17.5$ K. Note that $q \sim 0.75$ is a value commonly observed in other R$T_2$Al$_{10}$ compounds such as NdFe$_2$Al$_{10}$ and SmRu$_2$Al$_{10}$.\cite{Robert14,Takai15} 
The similar result for GdRu$_2$Al$_{10}$ seems to show that $q \sim 0.75$ is the fundamental propagation vector of the RKKY interaction in the R$T_2$Al$_{10}$ system. 
It is also noteworthy that the temperature dependence of the $q$ value is very similar to that of the order parameter. 
As shown in the inset of Fig.~\ref{fig:TdepQInt}(b), it is proportional to the square root of the intensity, reflecting the magnitude of the $4f$ magnetic moment. 
From this plot, we see that the $q$ value obtained from the extrapolation to zero intensity is $q_0=0.746$, which is not exactly the commensurate value of 0.75. 

\subsection{Cycloidal magnetic structure}
\subsubsection{Magnetic structure model}
There are two atomic positions of Gd at the $4c$ site of the $Cmcm$ space group: Gd-1 at $\mib{d}_1=(0, y, 1/4)$ and Gd-2 at $\mib{d}_2=(0, \bar{y}, 3/4)$, where $y=0.1266$.\cite{Thiede98,Sera13b} 
In the present single-$\mib{q}$ magnetic structure, the magnetic moments, $\mib{\mu}_{1,j}$ and $\mib{\mu}_{2,j}$ of Gd-1 and Gd-2, on the $j$th lattice point at $\mib{r}_j=(n_{1j}, n_{2j}, n_{3j})$ and $(n_{1j}+1/2, n_{2j}+1/2, n_{3j})$ ($n_{1j}, n_{2j}, n_{3j}$ are integers) are generally expressed as
\begin{align}
\mib{\mu}_{1,j} &= \mib{m}_1 e^{i\mib{q} \cdot \mib{r}_j} +  \mib{m}_1^* e^{-i\mib{q} \cdot \mib{r}_j}\,, \\
\mib{\mu}_{2,j} &= \mib{m}_2 e^{i\mib{q} \cdot \mib{r}_j} +  \mib{m}_2^* e^{-i\mib{q} \cdot \mib{r}_j}\,,
\end{align}
where $\mib{m}_1$ and $\mib{m}_2$ are the magnetic amplitude vectors of Gd-1 and Gd-2, respectively. 
In the present case of GdRu$_2$Al$_{10}$, since it is expected from the $\chi(T)$ and $M(H)$ behaviors that the moments are ordered in the $bc$ plane, 
it is also expected that $\mib{m}_1$ and $\mib{m}_2$ can be written using the $b$ and $c$ axis components as 
$\mib{m}_1 = m_1 (\hat{\mib{b}} + e^{i\varphi} \hat{\mib{c}})$ and $\mib{m}_2 = m_2 e^{i\alpha} (\hat{\mib{b}} + e^{i\varphi} \hat{\mib{c}})$, where $\hat{\mib{b}}$ and $\hat{\mib{c}}$ represent the unit vectors along the $b$ and $c$ axes, respectively, $\varphi$ the phase difference between the $b$ and $c$ axis components, $\alpha$ the phase difference between the moments of Gd-1 and Gd-2, and $m_1$ and $m_2$ the amplitudes of the moments of Gd-1 and Gd-2, respectively. 
These expressions for the magnetic structure are reduced to the following:
\begin{align}
\mib{\mu}_{1,j} &= m_1 \bigl\{ \hat{\mib{b}}  \cos \mib{q}\cdot\mib{r}_j  + \hat{\mib{c}}  \cos (\mib{q}\cdot\mib{r}_j + \varphi) \bigr\} \,, \label{eq:magst1} \\
\mib{\mu}_{2,j} &= m_2 \bigl\{ \hat{\mib{b}} \cos (\mib{q}\cdot\mib{r}_j + \alpha) + \hat{\mib{c}} \cos (\mib{q}\cdot\mib{r}_j + \varphi + \alpha) \bigr\}\;. \label{eq:magst2} 
\end{align}
Since $\mib{q}$ is along the $b$ axis, this structure generally represents a cycloidal magnetic structure. 
When $\varphi=\pi/2$, it describes a perfect cycloid in which the adjacent moments have a fixed angle. 
When $\alpha=\pi/2$, the moments of Gd-1 and Gd-2 at the nearest-neighbor positions are antiferromagnetically coupled. 

\begin{figure}[t]
\begin{center}
\includegraphics[width=8cm]{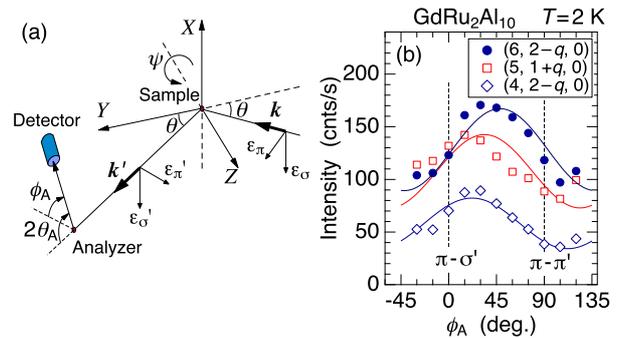}
\end{center}
\caption{(Color online) Scattering geometry of the experiment for polarization analysis. The Cu-220 reflection is used as an analyzer. 
The incident X-ray is $\pi$-polarized. $2\theta_{A}$ is $84.04^{\circ}$ at 7.246 keV. 
(b) $\phi_{A}$ dependence of the resonant intensity at 7.246 keV. Solid lines are the calculations assuming a cycloidal magnetic structure as described in the text. }
\label{fig:Polana1}
\end{figure}

\subsubsection{Polarization analysis}
To examine the magnetic structure, we have performed a polarization analysis measurement at 2 K for three peaks in the $(H, K, 0)$ plane. The scattering geometry and the results are shown in Fig.~\ref{fig:Polana1}. 
The solid lines are the calculations using the following expression for the $E1$-$E1$ scattering amplitude of magnetic dipole origin:
\begin{equation}
\hat{F}_{E1E1} \propto i(\mib{\varepsilon}' \times \mib{\varepsilon}) \cdot \sum_{n,j} \mib{\mu}_{n,j} e^{-i\mib{Q}\cdot(\mib{r}_j+\mib{d}_n)}\;,
\label{eq:FE1}
\end{equation}
where $n=1,2$ represents the Gd site and $\mib{Q}=\mib{k}'-\mib{k}$ is the scattering vector. 
Note that $\hat{F}_{E1E1}$ is expressed as the scalar product of the geometrical factor for the rank-1 $E1$-$E1$ resonance, 
$\mib{G}_{E1E1}=i(\mib{\varepsilon}' \times \mib{\varepsilon})$, and the magnetic structure factor (rank-1 tensor).\cite{Nagao05}
We assumed an equal amplitude of $m_1=m_2$, an antiferromagnetic coupling of $\alpha=\pi/2$, and a slightly modified cycloid of $\varphi=\pm 0.4\pi$. 
If we assume a perfect cycloid of $\varphi=\pm 0.5\pi$, we obtain almost flat $\phi_{A}$ dependence, which clearly disagrees with the data exhibiting a significant oscillation. 
If we assume a collinear structure with $\varphi=0$, the intensity should vanish at some $\phi_A$, which also disagrees with the data.  
From these results, we can conclude that the magnetic structure of GdRu$_2$Al$_{10}$ is a modified cycloid. 
Although the assumption of $\alpha=\pi/2$ is not validated here, it will be described later.

\subsubsection{Measurement using circularly polarized X-rays}
\begin{figure}[t]
\begin{center}
\includegraphics[width=7.5cm]{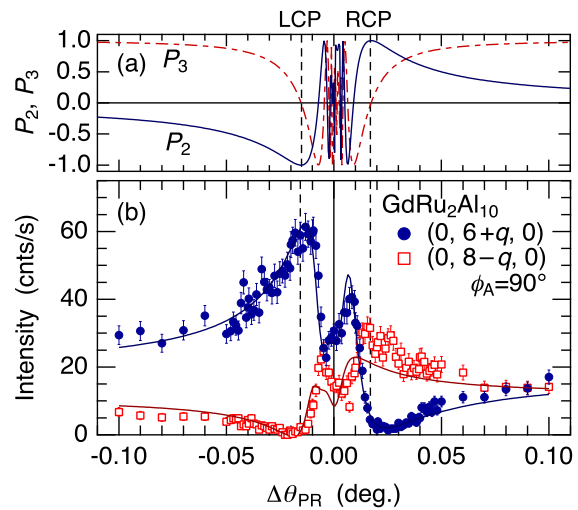}
\end{center}
\caption{(Color online) (a) $\Delta \theta_{\text{PR}}$ dependence of the incident Stokes parameters. 
The vertical dashed lines represent the positions of LCP and RCP states. The beam is depolarized around $\Delta \theta_{\text{PR}}=0$. 
(b) $\Delta \theta_{\text{PR}}$ dependence of the intensities of the $(0, 6+q, 0)$ and $(0, 8-q, 0)$ magnetic Bragg peaks. 
The analyzer is set at $\phi_{A}=90^{\circ}$. The $a$, $b$, and $c$ axes of the sample are set to coincide with the $Y$, $Z$, and $X$ axes, respectively, in Fig.~\ref{fig:Polana1}(a). }
\label{fig:PRscan}
\end{figure}

Another experimental result providing direct evidence for the cycloidal structure is shown in Fig.~\ref{fig:PRscan}. 
In this measurement, we inserted a diamond phase retarder in the incident beam and tuned the incident polarization state. 
By rotating the angle of the phase retarder $\theta_{\text{PR}}$ around the (1, 1, 1) Bragg angle $\theta_{\text{B}}$, where the scattering plane is tilted by $45^{\circ}$, a phase difference arises between the $\sigma$ and $\pi$ components in the transmitted beam, which is proportional to $1/(\theta_{\text{PR}} - \theta_{\text{B}})$. 
This allows us to tune the incident linear polarization to left-handed circular polarization (LCP) and right-handed circular polarization (RCP) by changing $\Delta \theta_{\text{PR}} =\theta_{\text{PR}} - \theta_{\text{B}}$. 
The polarization state of the incident beam as a function of $\Delta \theta_{\text{PR}}$ is shown in Fig.~\ref{fig:PRscan}(a) using the Stokes parameters $P_2$ ($+1$ for RCP and $-1$ for LCP) and $P_3$ ($+1$ for $\sigma$ and $-1$ for $\pi$). 
Since we used the vertical scattering plane configuration in this measurement 
[the configuration in Fig.~\ref{fig:Polana1}(a) is rotated by $90^{\circ}$ around the $Y$ axis], the incident linear polarization is $\sigma$ when $\Delta \theta_{\text{PR}}$ is large. 
$P_1$ ($+1$ for $+45^{\circ}$ and $-1$ for $-45^{\circ}$ linear polarization) is zero in this setup. 

As clearly demonstrated in Fig.~\ref{fig:PRscan}(b), the magnetic Bragg peaks at $(0, 6+q, 0)$ and $(0, 8-q, 0)$ exhibit the opposite dependence on $\Delta \theta_{\text{PR}}$. 
That is, the peak intensity at $\mib{\tau}+\mib{q}$ is strong for LCP but weak for RCP, whereas the relation becomes opposite at $\mib{\tau}-\mib{q}$. This is theoretically explained by assuming the cycloidal structure. 
For simplicity, if we assume a perfect cycloid with $\varphi=+\pi/2$ ($+$ helicity), the $E1$-$E1$ scattering amplitude matrix for a $(0, 2n \pm q, 0)$ peak, where $\mib{Q}$ is parallel to the propagation vector of the cycloid, is written as
\begin{equation}
\hat{F}_{E1E1} \propto \begin{pmatrix}
0 & \mp i \sin\theta \\ \mp i \sin\theta &  \sin 2\theta
\end{pmatrix}\;.
\end{equation}
Here, $\text{Re} F_{\pi\pi'} \cdot \text{Im} F_{\sigma\pi'}$ is responsible for the scattering cross section proportional to $P_2$. 
The intensity for an incident X-ray with $P_2=+1$ and with analyzer conditions of $2\theta_A=90^{\circ}$ and $\phi_A=90^{\circ}$ is calculated to be 
\begin{equation}
I \propto (1\mp 2\cos\theta)^2\sin\theta\,.
\end{equation}
When the helicity changes, or when $P_2=-1$, the $\mp$ and $\pm$ signs are interchanged. 
Although the expression is slightly modified for $\varphi\neq \pm\pi/2$, the basic mechanism of the asymmetric intensity is the same. 
The solid lines in Fig.~\ref{fig:PRscan}(b) are the calculated curves for the proposed magnetic structure of the modified cycloid. 

This experimental result shows that the helicity of the cycloid, i.e., the sign of $\varphi$, is uniquely determined in the sample without a formation of domains at least within the range of the beam size of $\sim 1\times 1$ mm$^2$ at the surface. 
Since the crystal space group $Cmcm$ has an inversion symmetry and mirror planes, the two cycloids with $+$ and $-$ helicity have equal energy and should form domains. If the two domains are mixed, the result in Fig.~\ref{fig:PRscan}(b) should exhibit a more symmetric curve. 
We consider that this is an accidental result caused by some surface strains due to polishing or by some external strain from the varnish used to glue the sample. 
Note that the following experimental results were also obtained in the single-domain state.
This provides us with a valuable opportunity to extract detailed information. 

\subsection{Local noncentrosymmetry and toroidal moment}
\begin{figure}[t]
\begin{center}
\includegraphics[width=7.5cm]{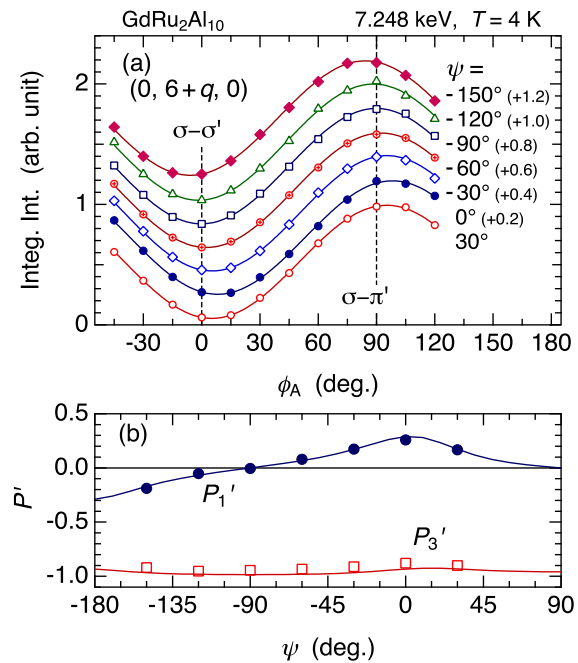}
\end{center}
\caption{(Color online) (a) $\phi_A$ scans at zero field and 4 K performed at several azimuthal angles. 
Integrated intensities were obtained by performing rocking scans of the analyzer crystal. 
The scattering plane is vertical and the incident polarization is $\sigma$. $\psi=0^{\circ}$ when the $a$ axis is in the scattering plane. 
(b) Azimuthal angle dependence of the Stokes parameters of the scattered X-ray, which is deduced from the $\phi_A$ scans in (a). }
\label{fig:AziscanPOL}
\end{figure}

Figure \ref{fig:AziscanPOL}(a) shows $\phi_A$ scans performed at several azimuthal angles using a four-circle diffractometer with a vertical scattering plane. 
The incident polarization is $\sigma$ in this configuration. 
The results in Fig.~\ref{fig:AziscanPOL}(a) show that the scattering is mostly $\sigma$-$\pi'$, which is a reasonable result for the $E1$-$E1$ resonance of magnetic dipole origin. However, the minimum and maximum positions of the intensity are slightly shifted from $0$ and $90^{\circ}$, respectively. If there is no $\sigma$-$\sigma'$ scattering, the intensity should have a minimum at $\phi_A=0^{\circ}$. 
This result shows that there is some $\sigma$-$\sigma'$ intensity, which is forbidden in the $E1$-$E1$ resonance of magnetic origin because $\mib{\varepsilon}_{\sigma}' \times \mib{\varepsilon}_{\sigma}=0$. 

\begin{figure}[t]
\begin{center}
\includegraphics[width=8cm]{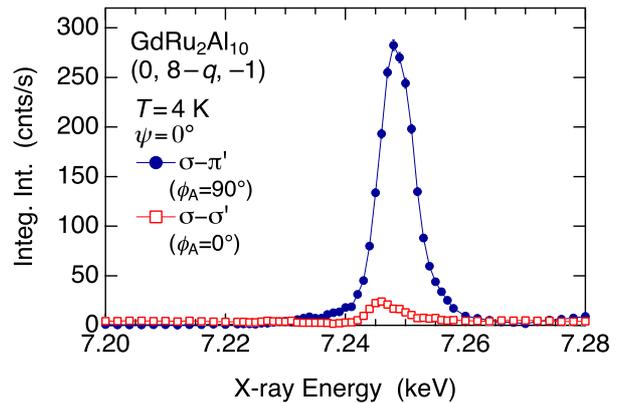}
\end{center}
\caption{(Color online) X-ray energy dependence of the $(0, 8-q, -1)$ peak intensity with polarization analysis. 
Data were taken in a vertical scattering plane configuration, where the incident polarization is $\sigma$.}
\label{fig:EscanSPSS}
\end{figure}

From the $\phi_A$ scans, we can extract the linear polarization states $P_1^{\,\prime}$ and $P_3^{\,\prime}$ of the scattered X-ray using the following expressions. 
The intensity at the detector after the analyzer crystal is expressed as 
\begin{equation}
I \propto \Bigl(\frac{d\sigma}{d\Omega}\Bigr) \Bigl\{1-\frac{1}{2}(1-P_{3A}) \sin^2 2\theta_A \Bigr\}\,,
\label{eq:int}
\end{equation}
where $(d\sigma / d\Omega)$ represents the scattering cross section at the sample and $P_{3A}$ is the Stokes parameter $P_3$ for the analyzer:
\begin{equation}
P_{3A} = -P_1^{\,\prime} \sin 2\phi_A + P_3^{\,\prime} \cos 2\phi_A \,.
\label{eq:P3A}
\end{equation}
The result of the analysis is shown in Fig.~\ref{fig:AziscanPOL}(b). 
This shows not only that $P_3^{\,\prime}$ is constantly close to $-1$, indicating that the scattering is mostly $\sigma$-$\pi'$ in the whole $\psi$ region, but also that $P_1^{\,\prime}$ exists, which is associated with the appearance of $\sigma$-$\sigma'$ scattering. 
The existence of $\sigma$-$\sigma'$ scattering is also directly demonstrated in Fig.~\ref{fig:EscanSPSS} by an energy scan. 

One possibility for the origin of the $\sigma$-$\sigma'$ scattering is an $E2$-$E2$ resonance through the $2p\leftrightarrow 4f$ transition. 
The scattering amplitude for an $E2$-$E2$ resonance of magnetic dipole origin (rank-1) is obtained by replacing the geometrical factor in Eq.~(\ref{eq:FE1}) with that of $E2$-$E2$: 
$\mib{G}_{E2E2}=i\{ (\mib{k}' \cdot \mib{k})(\mib{\varepsilon}' \times \mib{\varepsilon})
+(\mib{\varepsilon}' \cdot \mib{\varepsilon})(\mib{k}' \times \mib{k})
+(\mib{k}' \cdot \mib{\varepsilon})(\mib{\varepsilon}' \times \mib{k})
+(\mib{\varepsilon}' \cdot \mib{k})(\mib{k}' \times \mib{\varepsilon}) \}$.\cite{Nagao06}
By considering that the $E2$-$E2$ scattering interferes with the $E1$-$E1$ scattering, we can explain the appearance of $\sigma$-$\sigma'$ scattering and the $\psi$ dependence of $P_1^{\,\prime}$ and $P_3^{\,\prime}$ shown in Fig.~\ref{fig:AziscanPOL}(b). 
However, the energy position of the resonant peak for $\sigma$-$\sigma'$ is only 2 eV below that for $\sigma$-$\pi'$, suggesting that the resonance is mainly of $E1$ character.  
Since an $E2$-$E2$ resonance for the $L_3$ edge of a rare-earth atom typically occurs approximately 10 eV below the $E1$-$E1$ resonance, it is not likely that the $\sigma$-$\sigma'$ scattering is due to an $E2$-$E2$ resonance. 

Another possibility can be an $E1$-$E2$ resonance, which is normally forbidden for an atom located at a position with spatial inversion symmetry. In such cases, there is no mixing between $5d$ and $4f$. 
In the case of Gd sites in GdRu$_2$Al$_{10}$, the site symmetry $m2m$ lacks an inversion symmetry 
and the Gd ions are subjected to a finite electric dipole field $\mib{E}$, which is parallel to the $b$ axis and is oppositely oriented at Gd-1 and Gd-2. 
This allows an $E1$-$E2$ resonance through $5d$-$4f$ mixing. 
Since $\mib{E}$ is symmetrically equivalent to the position vector $\mib{r}$, it is also equivalently stated that there is a finite toroidal moment $\mib{r}\times\mib{\mu}$, a parity-odd rank-1 tensor, at the Gd sites.\cite{Lovesey05,Lovesey10} 
The structure factor for the toroidal moment in the present case can therefore be represented by 
$\sum_{n,j} (\mib{E}_n \times \mib{\mu}_{n,j}) e^{-i\mib{Q}\cdot(\mib{r}_j+\mib{d}_n)}$. 
This is finite at the same $\mib{q}$ vector as the magnetic cycloid; note that it is not ferrotoroidic.  
The geometrical factor of the $E1$-$E2$ resonance for the rank-1 tensor is expressed as 
\begin{equation}
\mib{G}_{E1E2} = (\mib{k}+\mib{k}') (\mib{\varepsilon'} \cdot \mib{\varepsilon}) + (\mib{k}' - \mib{k}) \times (\mib{\varepsilon}' \times \mib{\varepsilon})\,,
\end{equation}
and the scalar product with the structure factor gives the scattering amplitude. 
In Fig.~\ref{fig:AziscanPOL}, we show by solid lines the calculated Stokes parameters $P_1^{\,\prime}$ and $P_3^{\,\prime}$ by assuming that the toroidal moment is accompanied by the magnetic order, which well explains the data. 

\begin{figure}[t]
\begin{center}
\includegraphics[width=7.5cm]{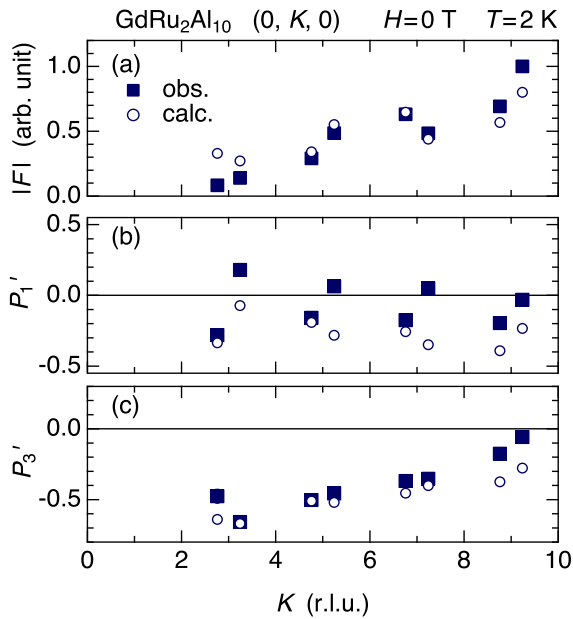}
\end{center}
\caption{(Color online) Scattering vector dependence of (a) the resonant scattering amplitude, 
(b) $P_1^{\,\prime}$, and (c) $P_3^{\,\prime}$. Squares and circles represent the observation and calculation, respectively. $|F_{\text{obs.}}|$ represents the square root of the scattering cross-section $\sqrt{(d\sigma/d\Omega)}$. The scattering plane is horizontal and the incident polarization is $\pi$. }
\label{fig:PolParams0T}
\end{figure}

In the horizontal scattering plane configuration, we performed the $\phi_{A}$ scans for several reflections of $(0, K, 0)$ along the $b$ axis. 
The parameters obtained,  the scattering amplitude $|F|\propto \sqrt{(d\sigma/d\Omega)}$, $P_1^{\,\prime}$, and $P_3^{\,\prime}$, using Eqs.~(\ref{eq:int}) and (\ref{eq:P3A}), are shown in Fig.~\ref{fig:PolParams0T}, where the 
observed values are compared with those of calculations assuming the modified cycloidal structure. 
The $E1$-$E2$ contribution is also taken into account to explain the result of $P_1^{\,\prime}$ for $(0, 6+q, 0)$. 
As shown in Fig.~\ref{fig:PolParams0T}(a), the calculation well explains the observed $|F|$. 
If the parameter $\alpha$ deviates from $\pm \pi/2$, the agreement becomes worse, indicating that the nearest-neighbor moments of Gd-1 and Gd-2 are antiferromagnetically coupled. The result of $P_3^{\,\prime}$ in (c) is also well explained by the present magnetic structure model. 
However, disagreement remains in the parameter $P_1^{\,\prime}$ as shown in (b). 
In the experiment, $P_1^{\,\prime}$ systematically oscillates between positive and negative, whereas in the calculation it is always negative. 
The relative amplitude and the phase parameter of the additional $E1$-$E2$ term with respect to the main $E1$ term were chosen so that the result for $(0, 6+q, 0)$ in Fig.~\ref{fig:AziscanPOL} was well reproduced. 
As a result, it seems that the $P_1^{\,\prime}$ data for the $(0, 2n+q, 0)$ reflections in Fig.~\ref{fig:PolParams0T}(b) are well explained, whereas those for $(0, 2n-q, 0)$ are not explained. 
Unfortunately, this disagreement cannot be improved in the present model. 

\subsection{Spin-flop transition in a magnetic field}
\begin{figure}[t]
\begin{center}
\includegraphics[width=7.5cm]{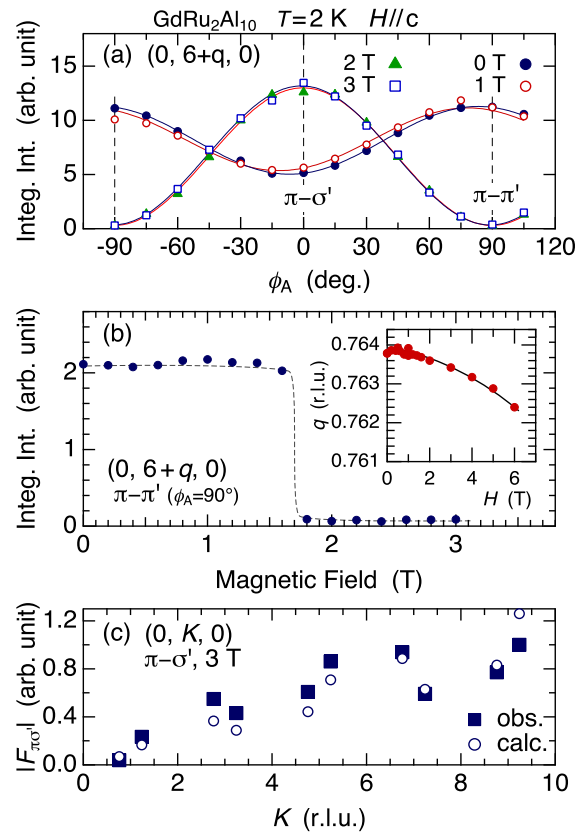}
\end{center}
\caption{(Color online) (a) $\phi_A$ scans at $(0, 6+q, 0)$ in magnetic fields along the $c$ axis. 
(b) Magnetic field dependence of the $\pi$-$\pi'$ intensity. The inset shows the field dependence of the $q$ value. 
(c) Scattering vector dependence of the resonant scattering amplitude for $\pi$-$\sigma'$ at 3 T. 
Observations are compared with the calculations. 
The scattering plane is horizontal and the incident polarization is $\pi$.
}
\label{fig:Polana3T}
\end{figure}

\begin{figure}[t]
\begin{center}
\includegraphics[width=8cm]{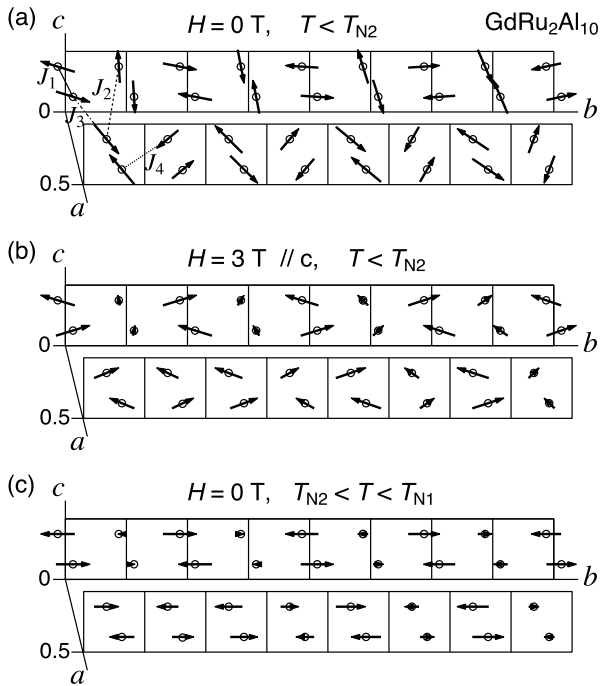}
\end{center}
\caption{Magnetic structure model of GdRu$_2$Al$_{10}$ at (a) 0 T and (b) 3 T along the $c$ axis, and (c) the intermediate phase $T_{\text{N2}} < T < T_{\text{N1}}$ . }
\label{fig:Magst}
\end{figure}

Figure \ref{fig:Polana3T}(a) shows the $\phi_A$ scans at $(0, 6+q, 0)$ in magnetic fields along the $c$ axis. 
At 0 and 1 T, the intensity does not vanish at any $\phi_A$, reflecting the cycloidal structure with both the $b$ and $c$ axis components. 
At 2  and 3 T above the transition field, the $\phi_A$ dependence markedly changes and the $\pi$-$\pi'$ intensity completely vanishes. 
The magnetic field dependence of the $\pi$-$\pi'$ intensity is shown in Fig. \ref{fig:Polana3T}(b). 
The disappearance of the $\pi$-$\pi'$ intensity above 1.7 T shows that the $c$ axis component vanishes. 
As shown in the inset, the $q$ value continuously and slightly decreases with increasing field. 
The scattering vector dependence of the scattering amplitude for $\pi$-$\sigma'$ is shown in Fig. \ref{fig:Polana3T}(c).
This result can be well explained by assuming a model structure with the $b$ axis component only, as shown by the open circles. 
If we include the $a$ axis component, the $|F_{\pi\sigma'}|$ data in Fig. \ref{fig:Polana3T}(c) cannot be reproduced. 
The magnetic structures at 0 and 3 T are summarized in Fig.~\ref{fig:Magst}. 

The transition in the magnetic structure from Fig.~\ref{fig:Magst}(a) to \ref{fig:Magst}(b) can be interpreted as a normal spin-flop transition, where the antiferromagnetic component parallel to the field vanishes and the moments become perpendicular to the field. 
They are canted to the field, giving rise to a ferromagnetic component. In Fig.~\ref{fig:Magst}(b), the ferromagnetic $c$ axis component is assumed uniform. 
Since the $b$ axis component is incommensurate, there are sites where the magnitude of the magnetic moment becomes small. 
However, such sites should have a larger $c$ component. If this is the case, it will give rise to higher harmonic intensities. 
The search for higher harmonics in magnetic fields, however, has not been performed.  

\subsection{Temperature-dependent cycloidal structure}
\begin{figure}[t]
\begin{center}
\includegraphics[width=8cm]{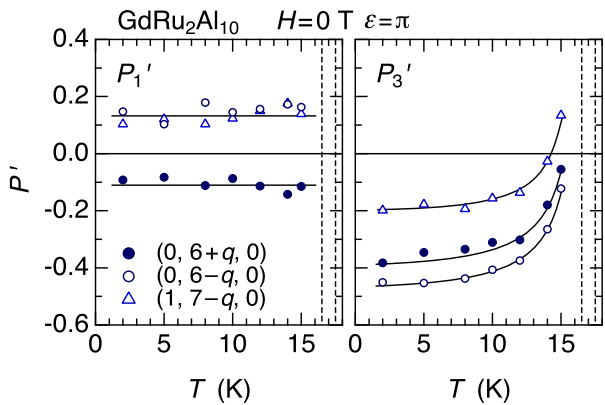}
\end{center}
\caption{(Color online) Temperature dependence of the Stokes parameters $P_1^{\,\prime}$ and $P_3^{\,\prime}$ for three reflections. 
Dashed vertical lines indicate the two transition temperatures. 
The scattering plane is horizontal and the incident polarization is $\pi$.}
\label{fig:TdepPolParam}
\end{figure}

Figure \ref{fig:TdepPolParam} shows the temperature dependence of the Stokes parameters $P_1^{\,\prime}$ and $P_3^{\,\prime}$, 
which was obtained from $\phi_A$ scans performed at several temperatures up to $T_{\text{N}}$. 
Although $P_1^{\,\prime}$ remains constant up to $T_{\text{N}}$, $P_3^{\,\prime}$ exhibits a strong temperature dependence on approaching $T_{\text{N}}$. 
If the magnetic structure remains cycloidal in the whole temperature range, as described by the model structure, these parameters should not change with the temperature. 
This result therefore shows that a significant change in the cycloidal structure takes place upon increasing the temperature to $T_{\text{N}}$. 

The temperature dependence of $P_3^{\,\prime}$ in Fig.~\ref{fig:TdepPolParam} can be interpreted as a change in the $c$-axis component. 
Observing in detail the magnetic susceptibility, only $\chi_b(T)$ shows a cusp at $T_{\text{N1}}=17.5$ K and $\chi_c(T)$ monotonically  increases down to $T_{\text{N2}}=16.5$ K. 
This shows that only the $b$-axis component is ordered below $T_{\text{N1}}$ and the $c$-axis component is ordered below $T_{\text{N2}}$. 
Without the $c$-axis component, which is perpendicular to the scattering plane here, the $\pi$-$\pi'$ scattering is forbidden and only the $\pi$-$\sigma'$ scattering takes place. In such a situation, $P_3^{\,\prime}=1$ should be realized. 
Although there are no data points above 15 K in Fig.~\ref{fig:TdepPolParam} because of the weak intensity, 
it seems that $P_3^{\,\prime}$ increases to unity on approaching $T_{\text{N2}}$. 
The temperature dependence of $P_3^{\,\prime}$ therefore reflects the ratio of the $c$-axis component to the $b$-axis component in the cycloid, which increases from nearly zero at $T_{\text{N2}}$ to unity at the lowest temperature. 
By comparison with the calculation, the ratio at 15 K is estimated to be $\sim 0.6$. 
The magnetic structure expected in the intermediate phase is shown in Fig.~\ref{fig:Magst}(c) by assuming that the $c$-axis component is zero. 

\section{Discussion}
In GdRu$_2$Al$_{10}$, the propagation vector changes with the temperature in proportion to the magnitude of the ordered moment. 
This shows that the RKKY interaction itself changes with the evolution of the ordered moment.\cite{JM91} 
The magnetic propagation vector reflects the $\mib{q}$ position where the exchange interaction $J(\mib{q})$, the Fourier transform of $J(\mib{r}_i - \mib{r}_j)$, is maximum. 
$J(\mib{q})$ for the RKKY interaction is associated with the local $c$-$f$ exchange interaction and $\chi(\mib{q})$ of the conduction electron system, where $\chi(\mib{q})$ is  determined by the form of the energy band structure. 
When a magnetic order develops on the Gd sites with a propagation vector $\mib{q}$, as described in Eqs.~(\ref{eq:magst1}) and (\ref{eq:magst2}), a perturbation of the exchange field to the conduction electron system arises, which is also described by the same $\mib{q}$ vector. 
As a result, a gap appears in the region of the Fermi surface where $\varepsilon_{\mib{k}'}=\varepsilon_{\mib{k}+\mib{q}}$ is satisfied.\cite{Watson68} 
This gap slightly modifies $\chi(\mib{q})$, and therefore the RKKY interaction $J(\mib{q})$ itself is also modified, resulting in a shift of the $\mib{q}$ vector.\cite{Elliot64} 
Then, the shift of the $\mib{q}$ vector from the original value of $\mib{q}_0$ just below $T_{\text{N}}$ becomes almost proportional to the ordered moment that develops. 
A similar temperature dependence of the $\mib{q}$ vector has also been reported in GdSi, GdNi$_2$B$_2$C, and GdPd$_2$Al$_3$.\cite{Feng13a,Feng13b,Detlefs96,Inami09}.

Note that when the magnetic anisotropy is taken into account, the temperature dependence of the $\mib{q}$ vector becomes more complicated, as observed in rare-earth metals.\cite{JM91} 
When there is a uniaxial anisotropy, a squaring up occurs and the third harmonic peak develops with decreasing temperature. 
This effect also causes the $\mib{q}$ vector to shift from $\mib{q}_0$. 
However, in this case, if we neglect the change in $J(\mib{q})$, the temperature dependence of the shift becomes proportional to $(T_{\text{N}} - T)^2$, which is different from the present case of Gd-based compounds.\cite{JM91,Sato94}

The cycloidal structure realized in GdRu$_2$Al$_{10}$ is associated with the very weak magnetic anisotropy of the $S=7/2$ state. 
In other isostructural R$T_2$Al$_{10}$ compounds the orientation of the magnetic moment is confined to be in a specific crystallographic axis because of the strong crystal field anisotropy and the nonvanishing orbital moment. 
As a result, the incommensurate magnetic structure with a propagation vector of $(0, q, 0)$ with $q\sim 0.75 $ to 0.8 necessarily induces a squaring up structure at low temperatures, giving rise to the appearance of the third harmonic component. 
In the case of GdRu$_2$Al$_{10}$ without such a uniaxial anisotropy, Gd spins can be oriented in any direction in the $bc$ plane. 
Combination with the maximum $J(\mib{q})$ at $(0, q\sim 0.75, 0)$ and the $bc$-plane anisotropy results in the cycloidal structure, which can release full magnetic entropy without forming a squaring up structure.  
Although it is very weak, however, an anisotropy exists with preferential ordering along the $b$ axis. 
This is actually observed in the magnetic structure just below $T_{\text{N1}}=17.5$ K, where the most intrinsic local anisotropy as well as the intrinsic $\mib{q}$ vector of the RKKY interaction can be observed because of the extremely small perturbation to the Fermi surface caused by the ordered moment. 
To obtain knowledge on the mechanism of the $b$-axis anisotropy, further analysis is required by considering the anisotropy of the conduction electrons due to the spin-orbit interaction, which interacts with the $4f$ electrons of Gd, as well as the dipole-dipole interaction between the Gd spins.\cite{Tosti03} 

The modification of the cycloidal structure with $\varphi=\pm 0.4\pi$ from the ideal one with $\varphi=\pm \pi/2$ is associated with the non-Bravais lattice of the Gd atoms. 
Since there is an intersite interaction between the magnetic moments at Gd-1 and Gd-2 sites, the perfect cycloid that should be realized in a single Bravais lattice is modified.

Finally, it is anomalous that only Ce$T_2$Al$_{10}$ among the R$T_2$Al$_{10}$ compounds has the magnetic propagation vector of $(0, 1, 0)$. The fact that all the magnetic structures of the R$T_2$Al$_{10}$ compounds reported to date have a $\mib{q}$-vector $(0, q, 0)$ with $q=0.75\!\sim\!0.8$ shows that the RKKY interaction of this system favors this $\mib{q}$ vector. 
Note, however, that it is not a simple problem to connect this common $\mib{q}$ vector to the Fermi surface structure reported for LaRu$_2$Al$_{10}$ since the Fermi surface does not seem to possess any particular nesting property.\cite{Sakoda11} 
In any case, the $(0, 1, 0)$ propagation vector of Ce$T_2$Al$_{10}$ is unusual and must be strongly associated with the Kondo semiconducting state caused by strong $c$-$f$ hybridization. 

\section{Conclusion}
We performed resonant X-ray diffraction experiments to clarify the magnetic structure of GdRu$_2$Al$_{10}$ 
and showed that a modified cycloidal structure propagating along the $b$ axis is realized in the $bc$ plane. 
The $\mib{q}$ vector shows a temperature dependence that is proportional to the magnetic order parameter, which was interpreted as being associated with partial gap formation in the conduction band and the resultant change in the magnetic exchange interaction. 
We also showed that the scattering involves not only the $E1$-$E1$ resonance of magnetic dipole origin but also the $E1$-forbidden $\sigma$-$\sigma'$ scattering. We interpreted this signal as being caused by the $E1$-$E2$ resonance due to the toroidal moments at the Gd sites, which is induced by the noncentrosymmetry and the magnetic order. 
The spin-flop transition for $H\parallel c$ was also studied and it was shown that the cycloidal structure changes to a canted incommensurate structure with the antiferromagnetic moments oriented along the $b$ axis. 
These properties were ascribed to the weak magnetic anisotropy of Gd ions with $S=7/2$, which show a slight preference for ordering along the $b$ axis, and the RKKY interaction mediated by the conduction electrons, which preferentially takes the $\mib{q}$ vector near $(0, 0.75, 0)$. 

\section*{Acknowledgements}
This work was supported by JSPS KAKENHI Grant Number 15K05175.
The synchrotron experiments were performed under the approval of the Photon Factory Program Advisory Committee (No. 2014G-129 and No. 2016G-159).

\end{document}